\documentclass[pre, showpacs, floatfix, preprint, endfloats*]{revtex4}

\usepackage{graphicx}
\usepackage{amsmath, amsfonts, amssymb, bm}
\usepackage[]{psfrag}

\psfragscanoff
\begin{document}

\title{Positronium in intense laser fields}
\author{Bj\"orn \surname{Henrich}}
\email{henrich@physik.uni-freiburg.de}
\author{Karen Z. \surname{Hatsagortsyan}}
\email{khats@physik.uni-freiburg.de}
\author{Christoph H. \surname{Keitel}}
\email{keitel@uni-freiburg.de}
\affiliation{Theoretische Quantendynamik, Physikalisches Institut, 
Universit\"at Freiburg, Hermann-Herder Str. 3, D-79104 Freiburg, Germany}

\date{\today}

\begin{abstract}
The dynamics and radiation of positronium is investigated in intense laser fields. 
 Our two-body quantum mechanical treatment displays the tunneling, free-evolution and recollision
dynamics of electron and positron both in the oscillating laser electric and laser magnetic field 
components. In spite of significant momentum transfer of the numerous incoming laser photons, 
recollisions of both particles are shown to occur automatically after tunneling ionization, along 
with substantial x-ray and gamma-ray emission during recombination and annihilation processes.
 \end{abstract}
 
\pacs{42.50.Hz, 42.65.Ky, 36.10.Dr, 78.70.Bj}

\maketitle

The highly nonlinear interaction of gaseous atoms with super intense laser pulses has been demonstrated
to give rise to the emission of coherent high-harmonic generation (HHG) up to the x-ray regime, with numerous 
applications in high-resolution spectroscopy and diagnostics (see e.g. \cite{hhg}).
Since the kinetic energy and consequently the frequency of the emitted radiation of a particle increases 
with rising laser intensity, considerable interest has been directed towards understanding the complex 
relativistic quantum dynamics of atoms and ions in ultra intense laser pulses \cite{Protopapas}. 
Once electrons reach velocities nonnegligible to that of light, however, the magnetically induced Lorentz 
force or in other words the momentum transfer of the numerous incoming photons induce a separation of 
electrons and ionic core in the laser propagation direction, resulting in a strong reduction 
of high harmonic yields \cite{Protopapas,Reiss,Walser}. While highly charged ions and crystals were
studied towards a reduction of this effect \cite{cry}, there is still a clear lack for an efficient system, where  
radiation pressure does not induce substantial ionization in the laser propagation direction and thus reduce coherent 
high frequency generation. On a different front there is also a quest for physics beyond atomic and classical 
plasma physics in ultra intense lasers, such as nuclear reactions \cite{ditmire} and QED effects \cite{qed}. 

\begin{figure}[b]
\includegraphics[width=12.8cm]{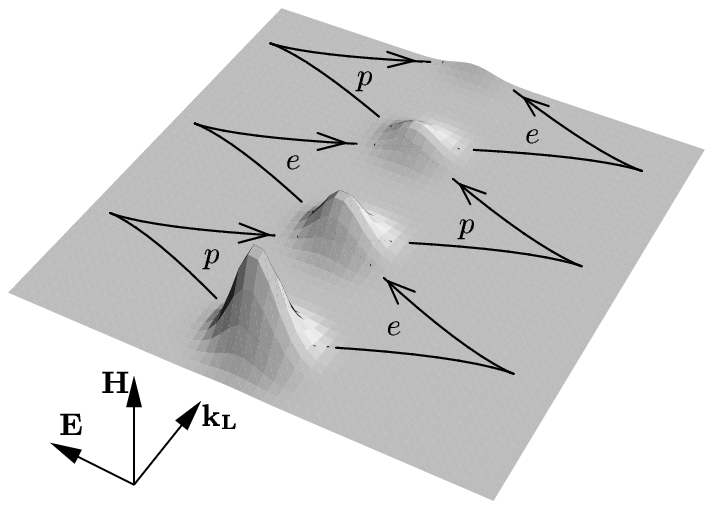}
\caption{\label{diagram}Schematic diagram displaying positronium dynamics in an 
intense laser field. The bound system depicted by the density of its 
wave function may ionize in the laser field. Once free, both electron $e$ and 
positron $p$ could be described as classical particles. Their trajectories are 
shown by the solid lines. The electric field ${\bf E}$ accelerates both particles in
opposite directions while the Lorentz force due to the magnetic field ${\bf H}$
leads to an identical drift in the propagation direction ${\bf k_L}$. Without an initial
center-of-mass motion these trajectories are symmetric and thus both particles
overlap periodically giving rise to possible recombinations and annihilations and thus
high-frequency light emission. 
} 
\end{figure}

In this letter the dynamics and high harmonic generation of positronium is investigated in intense laser fields.
The two-body system is shown to display unique properties: While tunnel-ionization of electron and positron may
occur almost oppositely in the laser polarisation direction, both particles sense the identical drift in the
laser propagation direction due to their equal magnitudes of mass and charge (see Fig. \ref{diagram}). 
Periodic electron-positron recollisions are shown to occur automatically head-on in spite of the influence of 
the Lorentz force. 
In addition to substantial coherent x-ray generation during periodic electron-positron recombinations we predict 
gamma radiation in the less likely events of laser-enhanced annihilations of both particles.  

Positronium consists of an electron and a positron and is known to be unstable. 
While ortho-positronium annihilates into three photons with a lifetime of 
$1.4\cdot 10^{-7}$s, para-positronium does so with two photons and a lifetime of
$1.25\cdot 10^{-10}$s \cite{Landau}. 
The presence of a strong laser field may induce substantial reductions \cite{Mittleman}
or enhancements \cite{Rivlin,Ritus} of the annihilation process into gamma photons.
However, even for the shorter lifetime of para-positronium it is sufficient for the interaction 
with many cycles of a femto second laser pulse with interaction lengths  not exceeding 
much the centimeter range \cite{Rivlin}. 

The laser field shall be described by the vector potential $\mathbf{A}(\mathbf{x},t)$ propagating in the z direction
and linearly polarized along the x direction. Being interested in the tunneling regime with 
moderately intense laser field strengths we restrict ourselves to Schr\"odinger dynamics in two
dimensions, though beyond the dipole approximation. The quantum dynamics of positronium in the laser field, 
taking fully into account of the Lorentz force due to the laser magnetic field component, 
is thus governed by the following Hamiltonian (in atomic units as throughout the article):
\begin{eqnarray}
H=\frac{\left(-i\nabla_{\mathbf{x}_e}-\frac{\mathbf{A}(t-z_e/c)}{c}\right)^2}{2}
 +\frac{\left(-i\nabla_{\mathbf{x}_p}+\frac{\mathbf{A}(t-z_p/c)}{c}\right)^2}{2}
  \nonumber \\
 -\frac{1}{|\mathbf{x}_e-\mathbf{x}_p|} \hspace{2.82cm}
    \label{e1} 
    \end{eqnarray}
where, $\mathbf{x}_i$ and $-i\nabla_{\mathbf{x}_i}$, ($i\in\{e,p\}$), 
are the operators of the coordinates and momenta of the electron and positron, respectively.
Further we introduce relative $\mathbf{r}=(x,y,z)=\mathbf{x}_e-\mathbf{x}_p $ and center of mass 
$\mathbf{R}=(X,Y,Z)=\frac{\mathbf{x}_e+\mathbf{x}_p}{2}$
coordinates. In accordance with the Ehrenfest theorem, the center of mass transversal canonical momenta 
$-i\partial_X$, $-i\partial_Y$ as well as $i\left(\partial_t+c\partial_Z\right)$ turn out to be 
conserved quantities, i.e. commute with $H$. The eigenvalues are defined to be $P_x, P_y, g$, respectively,
and  $\cal E$ is a separation constant that can be understood as the energy of the system before the interaction.
Therefore, we consider the ansatz 
\begin{equation}\Phi(\mathbf{r},\mathbf{R},\tau)=\exp\left(i\left(P_xX+P_yY-\frac{g}{c}Z-\cal
E\tau\right)\right)
  \phi(\mathbf{r},\tau) \label{e3} \end{equation}
singling out eigenfunctions of the conserved quantities in the wave function and introducing 
the running time $\tau=t-\frac{Z}{c}$.
Employing a $1/c$ expansion of the vector potential as function of the coordinates
of the relative motion of the particles, we obtain the following equation for $\phi$:
\begin{eqnarray}i\left(1-B_z\right)\partial_{\tau}\phi(\mathbf{r},\tau)= 
  \left\{\left(\frac{1}{i}\nabla_{\mathbf
  r}-\frac{\mathbf{A}(\tau)}{c}\right)^2+\frac{\mathbf{P}^2}{4}\right.\nonumber\\
  \left.+
  \frac{P_x\dot{A}(\tau)z}{2c^2} -\frac{1}{r}-{\cal E}-\frac{1}{4c^2}\partial^2_{\tau}
  \right\}\phi(\mathbf{r},\tau). \label{e4} \end{eqnarray}
  
Here, the center of mass velocity is introduced via $\mathbf{V}=\mathbf{P}/2$ with 
$\mathbf{B}=\mathbf{V}/c$ and $\mathbf{P}=\left(P_x,P_y,({\cal E} -g)/c\right)$.
The applied expansion takes into account of the leading multipole operators
describing the magnetically induced relative motion of electron and positron in the laser propagation
direction. Since $P_x$ is a conserved quantity,
this operator describes an oscillation of the relative coordinate in the
propagation direction while for atoms the analogous term would lead to a drift.
For consistency, the higher-order term for the center-of-mass motion $-\frac{1}{4c^2}\partial^2_{\tau}$
in Eq. (\ref{e4}) will be neglected in the following. 
 
We proceed by carrying out a transformation to the length gauge:
$\exp(-i\chi(\mathbf{r},\tau))\phi=\Psi$ with 
$\chi(\mathbf{r},\tau)=\mathbf{A}(\tau)\cdot\mathbf{r}/c.$
In this gauge we note the full velocity dependence of the transition matrix
elements  to be discussed later.
The resulting time dependent Schr\"odinger equation reads
\begin{eqnarray} & i\left(1-B_z\right)\partial_{\tau}\Psi(\mathbf{r},\tau)=
\bigg\{-\nabla^2_{\mathbf{r}} + I_p & \bigg.\nonumber\\
\bigg. & -E(\tau)\left(1-B_z\right)x-B_xE(\tau)z -\frac{1}{r} 
\bigg\}\Psi(\mathbf{r},\tau) &  
\label{e5} \end{eqnarray}

with electric field $\mathbf{E}(\tau)=-\frac{1}{c}\partial_{\tau}\mathbf{A}$ 
and positronium ionization potential $I_p$. 
Further we employ the reasonable assumptions proposed earlier by Lewenstein and
co-workers \cite{Lewenstein1} for atomic HHG in the dipole approximation, being: 
a) The contribution of all bound states except the ground state is neglected;
b) The depletion of the ground state is neglected;
c) In the continuum the electron is treated as a free particle in the laser field.
Then the time dependent wave function can be written as
\begin{equation}\left|\Psi(\tau)\right\rangle=
\left|0\right\rangle+\int_{-\infty}^{\infty} d^3\tilde{p}b(\tilde{\mathbf{p}},\tau)\left|\tilde{\mathbf{p}}\right\rangle
 \label{e6} \end{equation}
Here  $\left|0\right\rangle$  is the ground state
and $b(\tilde{\mathbf{p}},\tau)$ denotes the amplitude of the corresponding continuum state 
$\left|\tilde{\mathbf{p}}\right\rangle$. 
Integration of Eq. (\ref{e5}) with the ansatz (\ref{e6}) 
yields for the ionization amplitude $b$:
\begin{eqnarray}b(\tilde{\mathbf{p}},\tau)&=&\frac{-i}{1-B_z}\int_{-\infty}^{\tau}
\left\langle\mathbf{p}-\mathbf{A}(\tau')/c  
 -  \mathbf{\Lambda}(\tau')\right|H_I(\tau')\left|0\right\rangle  \nonumber\\
& &\times  
\exp\left(-i S(\mathbf{p},\tau,\tau')\right) d\tau', \label{e7} \end{eqnarray}
 where the quasiclassical action is
$S(\mathbf{p},\tau,\tau')=\int_{\tau'}^{\tau}d\tau''\left\{\left(\mathbf{p}-\mathbf{A}(\tau'')/c
 -\mathbf{\Lambda}(\tau'')\right)^2+I_p\right\}/(1-B_z)$.
 Here, the new variable
 $\mathbf{p}=\tilde{\mathbf{p}}+\mathbf{A}(\tau)/c+\mathbf{\Lambda}(\tau)$ is
 introduced with $\mathbf{\Lambda}(\tau)=\Lambda(\tau)\mathbf{e}_z=
B_xA(\tau)/\left[c\left(1-B_z\right)\right]\mathbf{e}_z$
  and $H_I(\tau)=-E(\tau)x(1-B_z)-B_xE(\tau)z$.  
In solving the equation for $b$ we have neglected all terms 
involving continuum-continuum transitions of no interest for the 
high-frequency part of harmonic generation as in \cite{Lewenstein1}.

We calculate the dipole moment $\mathbf{x}_d(\tau)=\left\langle\Psi(\tau)\right|\mathbf{r}
\left|\Psi(\tau)\right\rangle$ of which the Fourier transform $\mathbf{x}_d([\omega-\mathbf{k \cdot V}]/[1-B_z])$
yields the Doppler shifted harmonic spectrum, where $\omega$ and ${\bf k}$ denote the emitted radiation frequency 
and wave vector, respectively. 
In the tunneling regime the integrals can be obtained using the saddle point method, 
$\nabla_{\mathbf{p}}S\left(\mathbf{p},\tau,\tau'\right)=0$. Sustaining all momenta up to first order in $1/c$ yields:
$p_x(\tau,\tau')=\int_{\tau'}^{\tau}d\tau''A(\tau'')/{c(\tau-\tau')},p_y=0, p_z(\tau,\tau') = B_xp_x(\tau,\tau').$ 
 The matrix elements can be evaluated analytically employing the relation \cite{Bethe}
 $\mathbf{d}(\mathbf{p})=(d_x,d_y,d_z)=\left\langle\mathbf{p}\right|\mathbf{r}
\left|0\right\rangle=4 i\left(I_p\right)^{5/4} \mathbf{p}/(\pi \left(\mathbf{p}^2+I_p \right)^3)$
and performing the integration with respect to $\mathbf{p}$ gives
\begin{eqnarray}x_d(\tau)&=&\frac{1}{\sqrt{i}}
\int_{-\infty}^{\tau}d\tau'\left(\frac{\pi}{\tau-\tau'}\right)^{3/2}\nonumber\\
& & \times d_x^{\ast}\left(\mathbf{p}(\tau,\tau')-\mathbf{A}(\tau)/c - \mathbf{\Lambda}(\tau) \right)\nonumber\\
& & \times E(\tau')d_x\left(\mathbf{p}(\tau,\tau')-\mathbf{A}(\tau')/c - \mathbf{\Lambda}(\tau') \right)\nonumber\\
& & \times \exp\left(-iS(\mathbf{p}(\tau,\tau'),\tau,\tau')\right)+c.c.,
 \label{e8a}\end{eqnarray}
with classical action reading $S(\mathbf{p}(\tau,\tau'),\tau,\tau')=\int_{\tau'}^{\tau}d\tau''
\big[\left(  \mathbf{p}(\tau,\tau')-\mathbf{A}(\tau'')/c - \mathbf{\Lambda}(\tau'')   \right)^2+I_p\big]/
\left(1-B_z\right)$.
Taking into account the pole in the $\tau'$ integration finally results in
\begin{eqnarray}x_d(\tau) & = &
\frac{1}{\sqrt{i}}\sum_{\tau_0}^{}\left(\frac{\pi}{\tau-\tau_0}\right)^{3/2}\frac{I_p^{3/4}}{2E(\tau_0)}\nonumber\\
& & \times d_x^*\Big(p_x(\tau,\tau_0)-\frac{A(\tau)}{c}\Big)\nonumber\\
& & \times \exp\Big\{-\frac{2}{3}\frac{I_p^{3/2}}{E(\tau_0)}(1+B_z)\Big\}\nonumber\\
& & \times \exp \big\{-i S\big(\mathbf{p}(\tau,\tau_0),\tau,\tau_0\big) \big\}+c.c..
\label{e8}\end{eqnarray}
The birth times $\tau_0$ are determined, for given $\tau$, by the condition $p_x(\tau,\tau_0)=A(\tau_0)/c$.
The leading terms in $1/c$ have been sustained only in the phases because they play a far
less substantial role in the prefactor being identical to the dipole case. 

The dipole moment  in Eq. (\ref{e8}) displays essential differences in comparison with atomic HHG 
which shows especially in the real exponential factor in Eq. (\ref{e8}): 
$C:=\exp[- 2 I_p^{3/2} (1+B_z)/(3 E(\tau_0))]$. 
We stress that in the atomic HHG case \cite{Walser} 
there is an additional contribution exchanging  $I_p$ by $I_p^a + c^{2}\xi ^{4}/2$  
due to the drift in the laser propagation direction which can strongly reduce the amplitude 
and thus the efficiency of HHG. Here $\xi=\sqrt{2} A /c^2$ and  $I_p^a$ denotes the atomic ionization
potential. The expression $C$, without the factor $1+B_z$, coincides with 
the corresponding result within the dipole approximation for atoms.
This means that for positronium, the magnetically-induced drift in the laser propagation direction has no 
diminishing impact on HHG.

\begin{figure}[b]
\includegraphics[width=11.0cm]{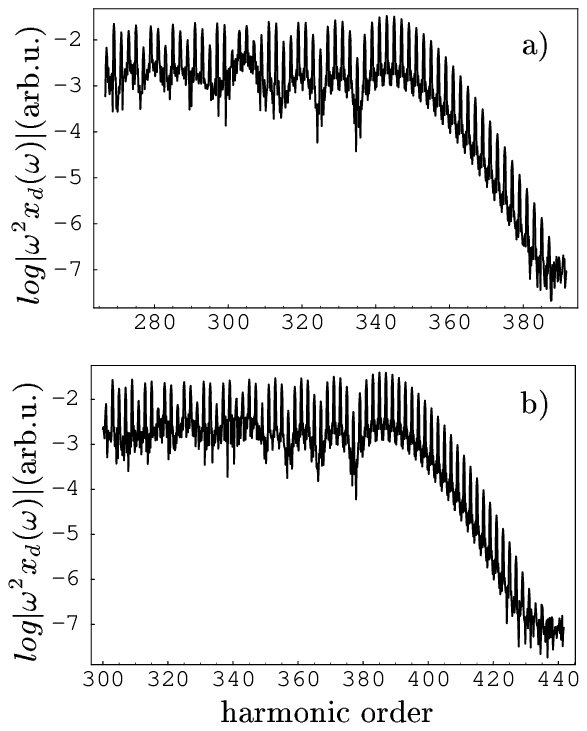}
\caption{\label{spektrum1}High harmonic spectra of positronium in the laser propagation direction 
in a laser pulse with vector potential $A=194.0$a.u. and angular frequency $\omega=0.0091$a.u. 
(corresponding to a laser intensity of 8.5 $\times 10^{12}$ W/cm${^2}$
and the second harmonic of a CO$_2$ laser with wavelenth 10 $\mu m$) and pulse shape parameters $c_1=1.0$ and $c_2=5.0$.
The symbols $c_1$ and $c_2$ denote the  number of cycles of ${\rm sin}^2$ turn-on/turn-off phases and of the intermediate cycles at full amplitude, respectively. The initial positronium velocity is
for a) $V_x=V_z=0.0001$a.u. and for b) $V_x=V_z=13.7$a.u. corresponding to 10$\%$ of the speed of 
light.
}  
\end{figure}

The figures \ref{spektrum1} display harmonic spectra  in the
laser propagation direction following the numerical integration of the dipole via Eq. (\ref{e8a}). 
In agreement with the numerical results, the maximal harmonic frequency $\omega_{max}$ is shown to critically 
depend on both the center-of-mass velocity and the observation direction via  the Doppler shift with 
$\omega_{max}(1-\mathbf{k \cdot V}/|\mathbf{k}|) = n_{max} \omega_L (1-B_z) = 3.17 U_p + I_p$ for
maximal harmonic number $n_{max}$ and  $U_p=c^{2}\xi ^{2}/2$.

In order to quantify the advantage of  positronium  for HHG in comparison to conventional HHG in atomic gas jets, 
we consider an example. In Lithium vapor being irradiated by a laser field at $\lambda =0.75 \mu $m wavelength 
and intensity $I=7\cdot 10^{16}$W/cm$^{2}$ ($\xi=0.17$), HHG up to the 10 keV photon energy 
range can be achieved in the tunneling regime for Li$^{2+}$ with $I_{p}^a=122.88$ eV.
Magnetic field effects in this case, however, are problematic as the extent of the electron drift during 
the laser period exceeds the atomic size: $\lambda \xi ^{2}/16\pi \gg a_B$, where $a_{B}$ is the Bohr radius. 
This substantially reduces the HHG efficiency, because of the earlier discussed $c^{2}\xi^4$ term being absent 
in positronium. Therefore, at $\xi =0.12$ (e.g. $\lambda \approx 50 \mu $m and 
$I=8\cdot 10^{12}$W/cm$^{2}$), the HHG cut-off frequency for positronium is equal to that of Li$^{2+}$ at the 
above mentioned parameters, however with a yield 15 orders of magnitude larger.
Thus, the low densities of positronium may be compensated for by the absence of drift-reduced recollisions.

Finally we discuss the possibility of  $\gamma $-ray emission with positronium via 
laser induced annihilation. In a super strong laser field with $\xi \gtrsim (mc^{2}/\hbar \omega )^{1/3} \approx 10^{2}$,
annihilation can occur via emission of $n \gtrsim 10^{6}$ photons \cite{Ritus} along with one $\gamma $-quantum (Fig. 3a). 
In more moderate laser intensities in the tunneling regime  with $\xi \leq 1$ of interest here, 
ortho-positronium, may annihilate into two $\gamma $-quanta (with induced emission of an odd number of laser photons, Fig. 3b)
 or spontaneously into three photons (Fig. 3c)).

We proceed with an order-of-magnitude estimation of $\gamma $-ray emission via strong laser 
assisted annihilation of ortho-positronium as depicted in Fig. 3b in comparison to spontaneous 3-photon annihilation (Fig. 3c).
The  2-vortex diagram in Fig. 3b, at first with bare particles, can be evaluated as product of the  Thomsonian 
cross-section $\sigma \approx\pi r_{0}^{2}$ with $r_{0}$ being the classical electron radius \cite{Landau} and the particle flux
to yield the annihilation probability $W\approx \sigma \cdot \rho c$, with electron and positron density $\rho =a^{-3}_B$.
The annihilation process is considered in a strong laser field here, with laser frequency being low in comparison
to the characteristic frequency $\omega_c=c^2$. In this situation the probability for bare particles may be multiplied by
a factor $F_s$ which takes into account of $s$ laser photons being emitted during the annihilation process (analogous to
Low theorem \cite{Low}).
The laser-induced differential probability of $\gamma $-ray emission at frequency $\omega_c$ 
being accompanied by the emission of $s$ laser photons can then be written as: 
$dW_{ind}\approx \pi r_{0}^{2}\rho c(d\omega /\Delta \omega _{D})(d\Omega /4\pi )\sum_s F_s$. 
Here we assumed that the $\gamma$-emission is distributed within a $4\pi $ solid angle $\Omega$ and that the Doppler
width $\Delta \omega _{D}$ describes the leading spectral broadening. Further 
$F_s=A_{0}^{2}(s\alpha \beta )+\xi ^{2}(2u-1)[A_{1}^{2}(s\alpha \beta )-A_{0}(s\alpha \beta )A_{2}(s\alpha \beta )]$ \cite{Ritus} 
is represented via generalized Bessel functions 
$A_{0}(s\alpha \beta ), A_{1}(s\alpha \beta ), A_{2}(s\alpha \beta )$, $\alpha =k^{'}A/(kq)c$, $\beta =-kk^{'}A^{2}/(8c^2(kq)^{2})$, 
$u=(kk^{'})^{2}/(4(kq)(kq^{'}))$, where $q_{\mu }=p_{\mu }-k_{\mu }A^{2}/(4kp)c^2$, $q^{'}_{\mu }=p^{'}_{\mu }-k_{\mu }A^{2}/(4kp)c^2$ 
for $\{\mu\}\in \{0,1,2,3\}$ and
$k, k^{'}, p, p^{'}$ denote the laser photon, $\gamma $-quanta, electron and positron 4-momenta, respectively. $A$ is the
4-vector-potential. In a weak laser field with $\xi \ll 1$ our formula for $dW_{ind}$ matches the known result in \cite{Rivlin}.
With a similar heuristic argumentation we may estimate the spontaneous 3-photon annihilation probability
without laser field at $\omega_c$ depicted in Fig. 3c, to give $dW_{sp}\approx \alpha \pi r_{0}^{2}\rho c(d\omega /\omega_c)(d\Omega /4\pi )$ 
with fine structure constant $\alpha=1/c$. This result for Fig. 3c , however, is well-known from the literature \cite{Landau}.

\begin{figure}[t]
\includegraphics[width=14cm]{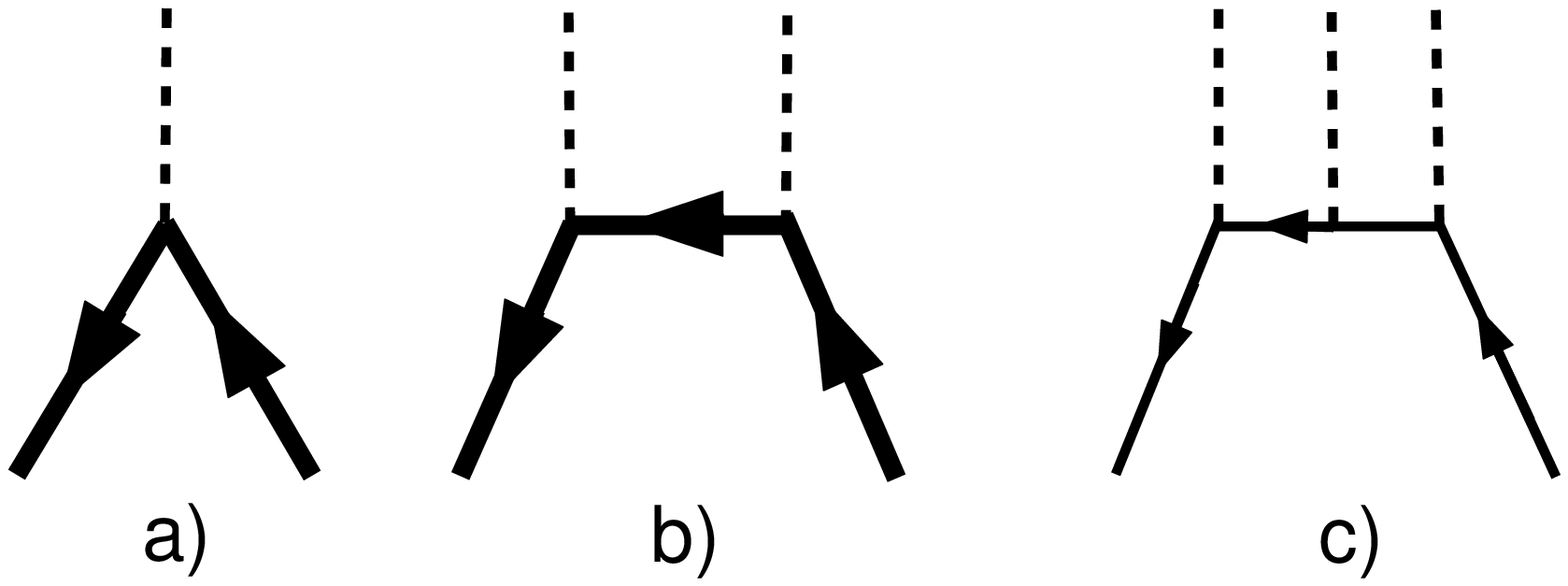}
\caption{\label{feynman} 
Feynman diagrams displaying $\gamma$-photon emission via electron-positron annihilation by 
a) multiphoton annihilation in super strong laser fields at $\xi \gtrsim 100$ with emission of 
$n\gtrsim 10^{6}$ laser photons along with one $\gamma $-quantum; 
b) annihilation in a laser field at $\xi \ll 100$ with emission of laser photons along with two
$\gamma $-quanta; 
c) annihilation of ortho-positronium with emission of three photons without laser field.
Bold lines correspond to the electron (positron) Volkov states, i.e. in the presence of the laser field,
dashed lines to the emitted  photons.}
\end{figure}


For moderately intense laser fields with $\xi \ll 1$, the main contribution to $dW_{ind}$ arises from 
$F_1\approx \xi ^{2}$, and at $\xi \approx 1$, for particular  $s$ there are arguments with
$F_s\approx 1$. In this latter case, the induced emission $dW_{ind}$ can exceed the spontaneous background $dW_{sp}$
up to a factor of  $(\omega_c/\Delta \omega _{D})/\alpha \approx 10^{6}$ for a Doppler linewidth 
$\Delta \omega _{D}/\omega_c\approx \sqrt(k_{B}T/\omega_c)\approx 10^{-4}$ with Boltzmann constant $k_{B}$
and temperature $T$. For para-positronium, however, a similar analysis shows that the intensity of 
$\gamma $-radiation via laser-induced annihilation can not exceed that via spontaneous annihilation. 
Regarding the $\gamma $-ray yield via laser-induced annihilation, a brilliance at photon energy $0.5$ MeV
of $B\simeq 10^{11}$ $photons/(s\cdot mm^{2}mrad^{2}(0.1\% bandwidth))$ can be estimated for 
 ortho-positronium density of $10^{10}$cm$^{-3}$ and an interaction length of $1$ cm.
This appears clearly detectable though moderate as compared to large-scale $\gamma $-ray
sources in the same spectral region
 \cite{Gizzi}.

Concluding, positronium was shown to provide various promising dynamical features in strong laser fields such as 
energetic head-on recollisions of its constituents, with substantial coherent x-ray generation and $\gamma $-ray emission.

This work has been supported by the German science foundation (Nachwuchsgruppe within SFB276) and an Alexander-von-Humboldt 
fellowship for KZH. 
We benefited from help on numerical issues by D.Diehl and A.Staudt, and especially from a seminal discussion 
with D. Habs which initiated this project.

\end{document}